\documentclass[preprint,12pt]{elsarticle}

\usepackage{graphicx}

\usepackage{makeidx}
\usepackage[]{hyperref}

\usepackage{amsmath}
\usepackage{amssymb,epsfig,graphics}
\usepackage{amscd}
\usepackage{verbatim}

\usepackage{amsmath}
\usepackage{amsthm}
\usepackage{amsfonts}

\usepackage{ulem}
\usepackage{color}

\begin{document}

\begin{frontmatter}

\title{Electronic states of zigzag graphene nanoribbons with edges reconstructed with topological defects}

\author[aff1,aff2]{R. Pincak\corref{cor1}}\ead{pincak@saske.sk}
\author[aff2,aff3]{J. Smotlacha\corref{cor1}}\ead{smota@centrum.cz}
\author[aff2]{V. A. Osipov}\ead{osipov@theor.jinr.ru}

\cortext[cor1]{Phone numbers: +7 925 1146637 (J. Smotlacha), +421 0907 936147 (R. Pincak)}

\address[aff1]{Institute of Experimental Physics, Slovak Academy of Sciences,
Watsonova 47,043 53 Kosice, Slovak Republic}
\address[aff2]{Bogoliubov Laboratory of Theoretical Physics, Joint
Institute for Nuclear Research, 141980 Dubna, Moscow region, Russia}
\address[aff3]{Faculty of Nuclear Sciences and Physical Engineering, Czech Technical University, Brehova 7, 110 00 Prague,
Czech Republic}

\date{\today}


\begin{abstract}
The energy spectrum and electronic density of states (DOS) of zigzag graphene nanoribbons with edges reconstructed with topological defects are investigated within the tight-binding method. In case of the Stone-Wales {\textit{zz}}(57) edge the low-energy spectrum is markedly changed in comparison to the pristine {\textit {zz}} edge. We found that the electronic DOS  at the Fermi level is different from zero  at any width of graphene nanoribbons. In contrast, for ribbons with heptagons only at one side and pentagons at another one the energy gap at the Fermi level is open and the DOS is equal to zero. The reason is the influence of uncompensated topological charges on the localized edge states, which are topological in nature. This behavior is similar to that found for  the structured external electric potentials along the edges.
\end{abstract}

\end{frontmatter}

\section{Introduction}Currently, there is a growing interest in studies of edge states in graphene structures. It has been found that zigzag graphene nanoribbons (ZGNRs) possess a localized edge state at the Fermi energy  which has a crucial influence on their electronic properties. In particular, the energy band gap of the ZGNRs is zero due to the existence of edge states and, consequently, these nanoribbons are always metallic. The presence of energy gap is necessary for various applications in nanoelectronics and, therefore, an important problem is to control and manipulate the edge states in ZGNRs.

Recently, the influence of external electric potentials applied along the edges of ZGNRs has been investigated. It was found that such potential can induce a spectral gap thus converting the metallic behavior of the ZGNR into a semiconducting one. As was mentioned in \cite{apelpal} this effect originates from the sensitivity of the spinorial edge states to electric potentials. What is interesting, the edge states are topological in nature~\cite{ryu,delplace}. Therefore one could expect a similar influence in case of topological charges situated along the edges. In order to check it we consider an artificial ZGNR with edges reconstructed with pentagons at one side and heptagons at the opposite side. For our motivation, it was shown in works \cite{EPJB1, EPJB2} that heptagonal defects influence the electronic structure of the graphene nanostructure significantly.

Our task is to study the electronic band structure of ZGNRs with reconstructed edges and to calculate the density of states (DOS). For this purpose, we employ
the well approved tight-binding method~\cite{wallace} which has been successfully used in studies of edge states in pristine ZGNRs~\cite{wakab}. The paper is organized as follows. In the next section, we give a brief description of the tight-binding method. Then, we study the energy band structure and the electronic DOS of endless ZGNRs containing periodically repeating structures with edges reconstructed with two different kinds of topological defects. A separate section is devoted to an analysis of the stability of the investigated structure by using the programs Avogadro \cite{avogadro} and GAMESS \cite{gamess}. Finally, we present a brief conclusion.

\section{Tight-binding method}
\begin{figure}[htbp]
{\includegraphics[width=20 pc]{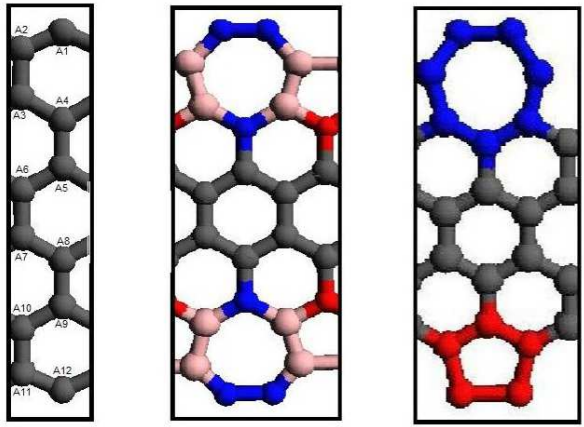}}\caption{The unit cells for pristine (left), {\textit{zz}}(57) (middle) and {\textit{zz}}(5/7) (right) zigzag graphene nanoribbons.}\label{unit}
\end{figure}

The tight-binding method assumes the numerical solution of  the stationary Schr\"{o}dinger equation
\begin{equation}\label{first}H\psi=E\psi,\end{equation}
where the Hamiltonian is written for $\pi$ electrons in graphene lattice with nearest-neighbors taken into account, $\psi$ is the linear combination of the wave functions which correspond to the particular atoms in the unit cell. In graphene lattice, the unit cell contains exactly two atoms while in graphene nanoribbons this number is much larger (see fig. \ref{unit}).
Let us enumerate them as $A_1, A_2,..., A_N$, where $N$ is the total number of atoms in the unit cell. Then
\begin{equation}\psi=C_1\psi_{A_1}+...+C_N\psi_{A_N},\end{equation}
and one can define the matrix coefficients
\begin{equation}H_{ij}=\int\psi_i^*H\psi_j{\rm d}\vec{r},\end{equation}
where $i,j\in\{A_1,...,A_N\}$. Owing to orthogonality of $\psi_i$ one gets
\begin{equation}\label{eq4}\sum\limits_{j=1}^NC_jH_{ij}=C_iES,\end{equation}
where the normalization condition is chosen to be $S=\int\psi_i^*\psi_i{\rm d}\vec{r}$ with $S$ being the number of unit cells in the nanostructure. Finally,  solving the matrix equation (\ref{eq4})
we obtain the energy eigenvalues and thereby the electronic spectrum of the given nanostructure. The electronic DOS is written as
\begin{equation}DOS(E)=\int\limits_0^{2\pi}\delta(E-E(\vec{k})){\rm d}\vec{k}.\end{equation}

\begin{figure}[htbp]
{\includegraphics[width=165mm]{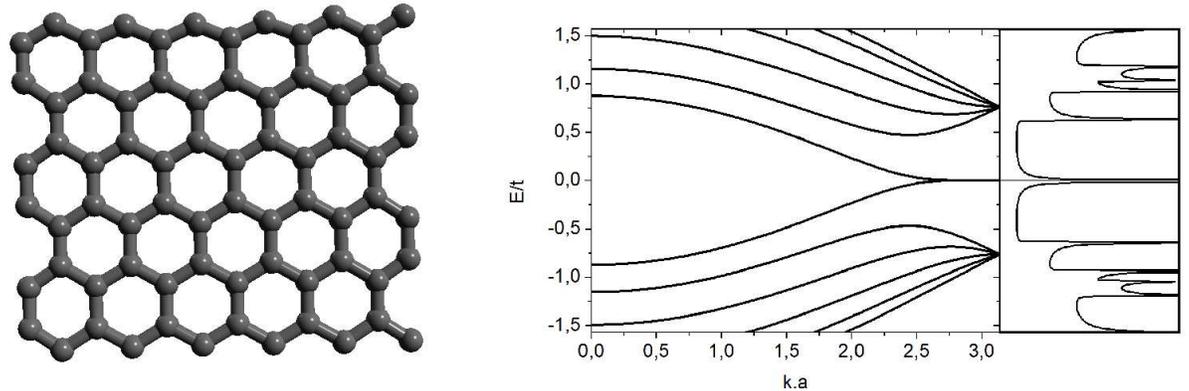}}\caption{Lattice structure, electronic spectrum and density of states of endless zigzag graphene nanoribbon. The results are in  perfect agreement with \cite{wakab}.}\label{fg1}
\end{figure}

In order to verify this consideration we have performed the numerical calculations of the electronic spectrum and the DOS for the ZGNR of given width. The results are shown in
fig. \ref{fg1} and they are in perfect agreement with \cite{wakab}.

\section{Zigzag graphene nanoribbons with reconstructed edges}Let us first consider ZGNRs with edges totally reconstructed with Stone-Wales defects in the form of repeated heptagon-pentagon pairs (also known as {\textit {zz}}(57) edge). The effect of such edge reconstruction on both the electronic band structure and the characteristics of low-energy edge states has been studied in detail in \cite{stonewales}. Evidence for graphene edges involving the Stone-Wales defects was given in \cite{koskinen}. For our purpose, we reconsider this problem within our approach and add an analysis of the electronic DOS. One has to take into account two important features: (i) this defect causes a deformation of the bond lengths  and corresponding angles between carbon atoms in heptagons and  pentagons and (ii) the energy of the C-C bonds is also changed.

\begin{figure}[htbp]
{\includegraphics[width=140mm]{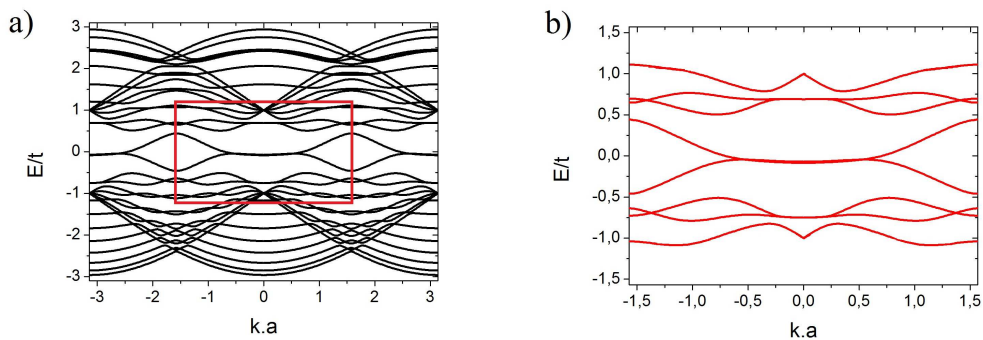}}\caption{Energy band structure of the zigzag graphene nanoribbon with {\textit {zz}}(57) edge (a). The low-energy region is highlighted in (a) and shown separately in (b).}\label{SW8c}
\end{figure}

We have calculated the complete electronic spectrum in the case when the edges formed by the Stone-Wales defects are separated by five lines of hexagons. The result is shown in fig. \ref{SW8c}(a). It should be mentioned that the low-energy part of the electronic spectrum highlighted in fig. \ref{SW8c}(a) and shown separately in fig. \ref{SW8c}(b) agrees well with the results in \cite{stonewales}.

\begin{figure}[htbp]
{\includegraphics[width=165mm]{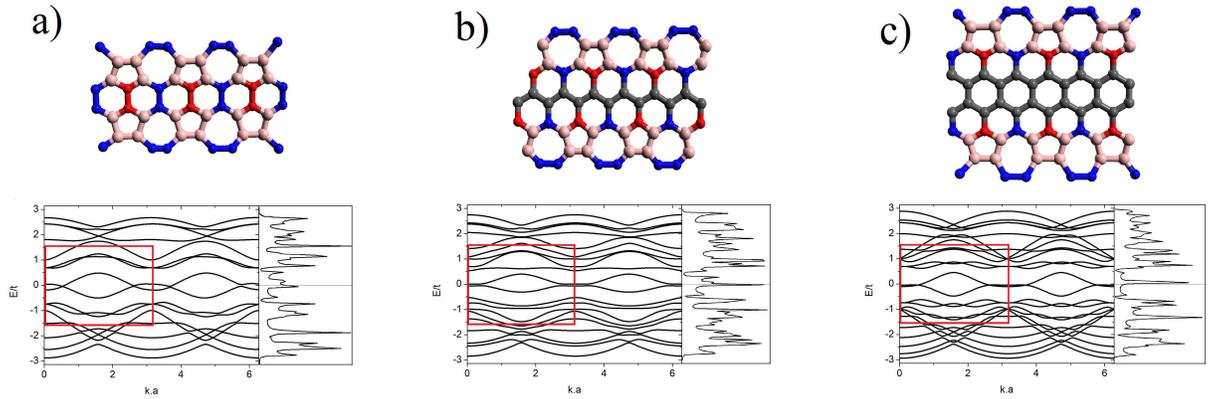}}\caption{Lattice structure, electronic spectrum and density of states of endless zigzag nanoribbons with {\textit{zz}}(57) edge. The low-energy region is highlighted.}\label{fg3}
\end{figure}

Fig. \ref{fg3} shows the electronic spectrum and the DOS for Stone-Wales reconstructed ZGNRs of different width.
As is seen, the DOS at the Fermi level decreases with a width.
An important conclusion, however, is that it remains finite even for very narrow ZGNRs.
It is interesting to note that band crossings occur at some wavevectors and the existing oscillations are incommensurate. Unfortunately, the origin of this phenomenon is not clear yet.
The velocity at the crossing decreases with increasing width of the ribbon.

\begin{figure}[htbp]
{\includegraphics[width=165mm]{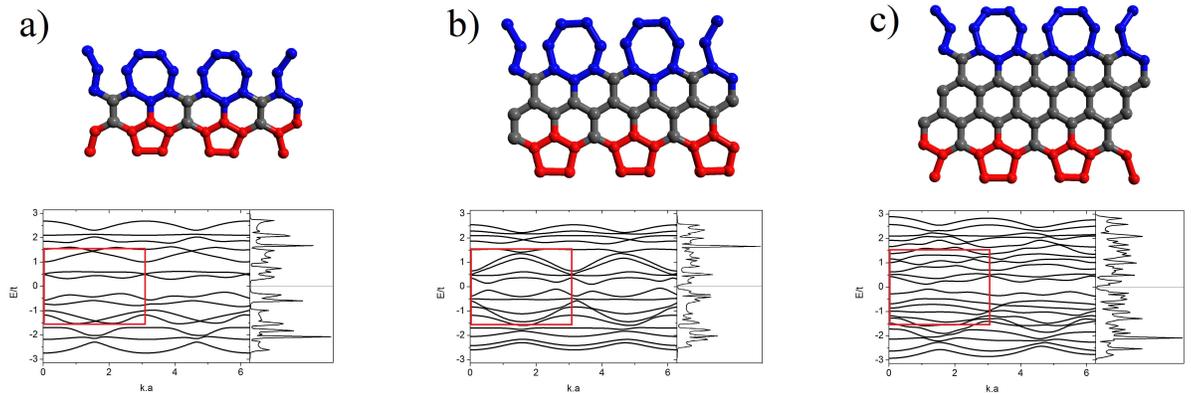}}\caption{Lattice structure, low-energy electronic spectrum and density of states of endless zigzag nanoribbons with {\textit {zz}}(5/7) edges. The low-energy region is highlighted.}\label{fg4}
\end{figure}

Finally, let us consider a case when heptagons and pentagons are situated along different edges of ZGNRs (see fig. \ref{fg4}). Let us call this configuration as {\textit {zz}}(5/7) edges. Notice that there is a principal distinction from the previous case. Indeed, the topological charges turn out to be compensated along the {\textit {zz}}(57) edges while here they have certain opposite signs at each side. This situation resembles the case with electrostatic potentials along the edges of ZGNRs considered in \cite{apelpal}.

The results of our calculations are shown in fig. \ref{fg4}. As is seen, the edge energies are markedly modified: a spectral gap is opened and the DOS at the Fermi energy turns out to be zero. The gap decreases with increasing width of the nanoribbon. Notice that this behavior fully complies with that found for external electric potentials in \cite{apelpal}. The authors \cite{apelpal} found an opening of wider gap for bigger electric potential at fixed width of ZGNR. In our case, we vary the width of ZGNRs at fixed topological charges.
Notice that in the case of the $zz$(5/7) edges (Fig. \ref{fg4}) the number of  band crossings falls significantly due to existing gap at the Fermi energy and a tendency of the spectral curves to lower their slope.

\section{Stability analysis of the ZGNR with {\textit {zz}}(5/7) edges}

\begin{figure}[htbp]
{\includegraphics[width=105mm]{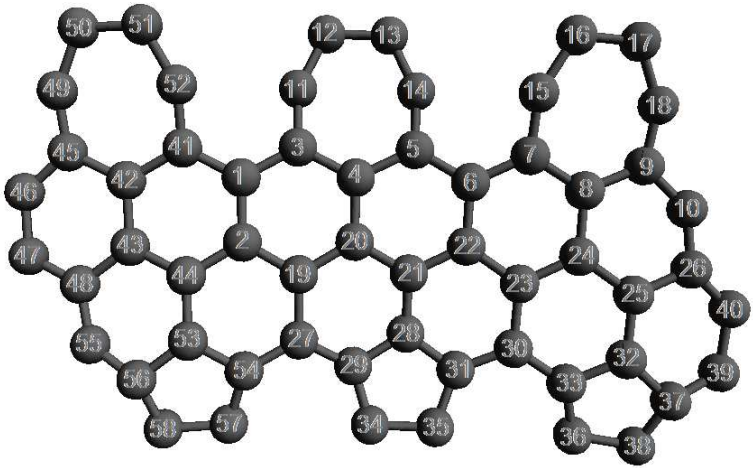}}\caption{A test molecule with enumerated atoms.}\label{fg5}
\end{figure}

The question arises of whether the artificial nanoribbon presented in fig. \ref{fg4} is stable. Indeed, additional bonds could be formed between some nearest carbon atoms at the edges thus transforming this configuration into the structure which would be more similar to that in fig. \ref{fg3}. In order to clarify this question we consider a bit simpler structure shown in fig. \ref{fg5} where we enumerate the individual atoms. The geometry of this molecule was optimized with the help of the program Avogadro. Then, we determine the wave functions for each atom and the multiplicity of the bonds between the atoms by using the program GAMESS designed to the quantum chemical computing. When the multiplicity is much lower than unity or equal to zero one can ignore the potential bond between a concrete pair of atoms.

The corresponding part of the output file of the program GAMESS is shown in  Table \ref{table}. It is represented by 3 columns, each consisting of 3 subcolumns. In each line, the first subcolumn indicates the numbers of atoms from the investigated pair as labeled in fig. \ref{fg5}. The second subcolumn gives the distance between these two atoms in the molecule and the multiplicity of the potential bond between these atoms is written in the third subcolumn.

We are interested in the bond order between atoms which could significantly change the structure of the nanoribbon's edge, i.e. the interesting pairs of atoms are 14-15, 11-52, 34-57 and 35-36 (see fig. \ref{fg5}). The relevant lines in Table \ref{table} are marked in bold. We see that for all of these pairs the corresponding bond orders take a negligible value. Furthermore, the pair 35-36 is not presented in the Table because the corresponding bond order does not exceed the chosen threshold 0.05 (see the first line in Table \ref{table}). This analysis shows that the investigated structure is stable. For wider nanoribbons like those in fig. \ref{fg4}, the distance between the atoms in the investigated pairs may be reduced and the corresponding bond order can be significant. This could lead to the modification of the structure.

\begin{figure}[htbp]
{\includegraphics[width=105mm]{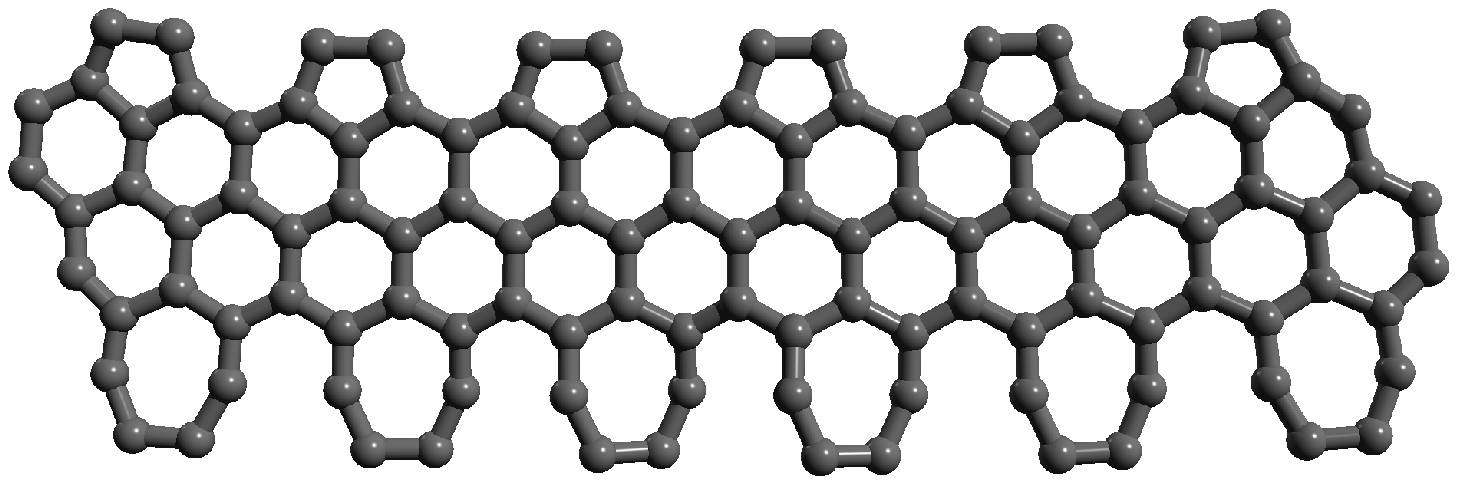}}
{\includegraphics[width=150mm]{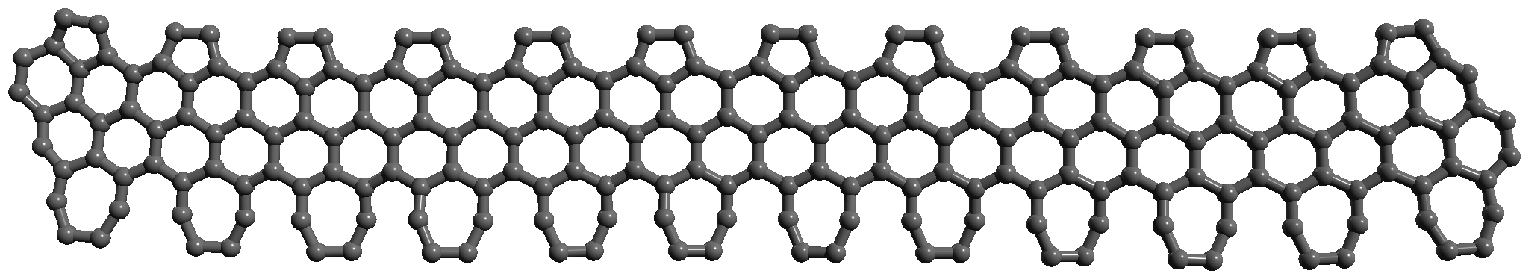}}\caption{Structures tested with the help of the program KVAZAR.}\label{fg6}
\end{figure}

This possibility was verified with the help of the molecular-mechanical method within the program KVAZAR \cite{kvazar}. The shorter molecule in Fig. \ref{fg6} was tested for the time interval 20\,ps and the longer molecule for the time interval 40\,ps. In the first case, all the critical bonds finally appeared while for the longer molecule only half of them were formed. This means that the fabrication of stable structures may require
the use of supporting experimental procedures such as decreasing temperature, adding the hydrogen atoms or some other substrates, etc.

\renewcommand{\baselinestretch}{1.0}

\begin{table}[htbp]
\caption{The bond orders between the carbon atoms of the structure in Fig. \ref{fg5}.}\label{table}
\resizebox{\textwidth}{!}{
\begin{tabular}{ccccccccccc}
\multicolumn{11}{c}{BOND ORDER AND VALENCE ANALYSIS \hspace{1cm} BOND ORDER THRESHOLD=0.050}\\
\multicolumn{11}{c}{- - - - - - - - - - - - - - - - - - - - - - - - - - - - - - - \hspace{1cm}
  - - - - - - - - - - - - - - - - - - - - - - - - - -}\\\\

            &        & BOND  & &            &        &  BOND  & &           &        &  BOND\\
  ATOM PAIR & DIST   & ORDER & &  ATOM PAIR & DIST   &  ORDER & & ATOM PAIR & DIST   &  ORDER\\
    1  2    & 1.495  & 1.245 & &     1   3  & 1.493  &  1.161 & &   1  41   & 1.498  &  1.148\\
     ...    &  ...   &  ...  & &      ...   &  ...   &   ...  & &    ...    &  ...   &   ... \\
     ...    &  ...   &  ...  & &      ...   &  ...   &   ...  & &    ...    &  ...   &   ... \\
     ...    &  ...   &  ...  & &      ...   &  ...   &   ...  & &    ...    &  ...   &   ... \\
   10  39   & 3.947  & 0.056 & &    11  12  & 1.522  &  1.472 & &  11  13   & 2.555  &  0.307\\
   11  14   & 2.797  & 0.080 & &    11  18  & 8.360  &  0.054 & &  \textbf{11  52}   & \textbf{2.823}  &  \textbf{0.054}\\
   12  13   & 1.504  & 1.642 & &    12  14  & 2.553  &  0.190 & &  13  14   & 1.522  &  1.470\\
   \textbf{14  15}   & \textbf{2.826}  & \textbf{0.197} & &    14  18  & 5.576  &  0.101 & &  15  16   & 1.525  &  1.776\\
   15  17   & 2.596  & 0.085 & &    15  18  & 2.854  &  0.196 & &  16  17   & 1.508  &  1.133\\
     ...    &  ...   &  ...  & &      ...   &  ...   &   ...  & &    ...    &  ...   &   ... \\
     ...    &  ...   &  ...  & &      ...   &  ...   &   ...  & &    ...    &  ...   &   ... \\
     ...    &  ...   &  ...  & &      ...   &  ...   &   ...  & &    ...    &  ...   &   ... \\
   33  36   & 1.421  & 1.380 & &    33  38  & 2.422  &  0.073 & &  33  39   & 3.682  &  0.051\\
   34  35   & 1.532  & 1.521 & &    \textbf{34  57}  & \textbf{3.317}  &  \textbf{0.301} & &  34  58   & 4.842  &  0.126\\
   35  57   & 4.848  & 0.058 & &    36  38  & 1.531  &  1.803 & &  36  40   & 4.725  &  0.083\\
     ...    &  ...   &  ...  & &      ...   &  ...   &   ...  & &    ...    &  ...   &   ... \\
     ...    &  ...   &  ...  & &      ...   &  ...   &   ...  & &    ...    &  ...   &   ... \\
     ...    &  ...   &  ...  & &      ...   &  ...   &   ...  & &    ...    &  ...   &   ... \\
   57  58   & 1.529  & 1.463 & &            &        &        & &           &        &
\end{tabular}
}
\end{table}

\renewcommand{\baselinestretch}{1.5}

\section{Conclusion}In conclusion, we have studied the effect of topological charges situated along the edges of ZGNRs on the edge state. For this purpose, we have numerically calculated the energy spectrum and the electronic DOS within the tight-binding method. The edge state is found to be noticeably modified in case of compensated charges for {\textit {zz}}(57) edge. Nevertheless, the DOS at the Fermi level remains finite at any width thus supporting the metallic behavior of this type of ZGNRs. The situation changes markedly for uncompensated topological charges. The nonzero energy gap is found to be induced which decreases with increasing width of the ZGNR with {\textit {zz}}(5/7) edges. This behavior agrees with the results obtained for decreasing electric potential in \cite{apelpal}. Similarly to the case of the external electric potential, this finding has a clear physical explanation coming from the topological origin of the zero-energy edge state in ZGNRs. Notice that the considered artificial ZGNR includes not only a non-trivial topology but excess electrical potentials as well. Indeed, there are additional charges at the five- and seven-membered rings placed into the pristine graphene lattice, which were estimated as -0.07 and +0.04, respectively~\cite{tamura}. In the case of the ZGNR with {\textit {zz}}(5/7) edges, the calculated difference of charges at both sides takes the value of 0.02 which roughly corresponds to the external electrostatic potential equal to $\sim$ $0.004$ eV at ribbon's width $\sim$ $0.7$ nm. This potential is too small to induce any observable spectral gap and, therefore, we conclude that namely the non-trivial topology and uncompensated topological charges play a crucial role in the gap opening.

It is interesting to note that a similar effect of band crossings appears in the linear oligoacenes \cite{signature} which are formed by fused benzene rings. It was found that close to certain crossings, the LUMO and HOMO interchange and, in this situation, the gap has a minimum. The local maxima of the gap manifest another level crossing. The observed oscillations in fundamental excitation gaps with increasing length of the oligoacene may be of practical interest in organic electronics and photovoltaics. We expect that our findings may also have prospective technological applications in graphene-based electronics.

\section*{Acknowledgments}
The authors acknowledge V.L. Katkov and O.E. Glukhova for useful discussions. Our special thanks to O.E. Glukhova and her research group for performing calculations by using the program KVAZAR.
This work has been supported by the VEGA Grant No. 2/0037/13 and by the Russian Foundation for Basic Research under Grant No. 12-02-01081. R. Pincak would like to thank the TH division in CERN for hospitality.

\end{document}